\documentclass[aps,preprint]{revtex4}
\usepackage{amssymb,epsf}
\usepackage{amsmath}
\usepackage{graphicx}

\begin{document}

\title{Thermodynamics of Rotating Charged Black Branes in
Third Order Lovelock Gravity and the Counterterm Method}
\author{M. H. Dehghani$^{1,2}$\footnote{email address:
mhd@shirazu.ac.ir} and R. B. Mann$^{3,4}$\footnote{email address:
rbmann@sciborg.uwaterloo.ca}}

\affiliation{$^1$Physics Department and Biruni Observatory,
College of Sciences, Shiraz
University, Shiraz 71454, Iran\\
$^2$Research Institute for Astrophysics
and Astronomy of Maragha (RIAAM), Maragha, Iran\\
$^3$Department of Physics, University of Waterloo, 200 University
Avenue West, Waterloo, Ontario,
Canada, N2L 3G1 \\
$^4$Perimeter Institute for Theoretical Physics, 35 Caroline St.
N., Waterloo, Ont. Canada}

\begin{abstract}
We generalize the quasilocal definition of the stress energy
tensor of Einstein gravity to the case of third order Lovelock
gravity, by introducing the surface terms that make the action
well-defined. We also introduce the boundary counterterm that
removes the divergences of the action and the conserved quantities
of the solutions of third order Lovelock gravity with zero
curvature boundary at constant $t$ and $r$. Then, we compute the
charged rotating solutions of this theory in $n+1$ dimensions with
a complete set of allowed rotation parameters. These charged
rotating solutions present black hole solutions with two inner and
outer event horizons, extreme black holes or naked singularities
provided the parameters of the solutions are chosen suitable. We
compute temperature, entropy, charge, electric potential, mass and
angular momenta of the black hole solutions, and find that these
quantities satisfy the first law of thermodynamics. We find a
Smarr-type formula and perform a stability analysis by computing
the heat capacity and the determinant of Hessian matrix of mass
with respect to its thermodynamic variables in both the canonical
and the grand-canonical ensembles, and show that the system is
thermally stable. This is commensurate with the fact that there is
no Hawking-Page phase transition for black objects with zero
curvature horizon.
\end{abstract}
\maketitle

\section{Introduction}

In four dimensions, the Einstein tensor is the only conserved symmetric
tensor that depends on the metric and its derivatives up to second order.
However for spacetimes possessing more than four dimensions, as assumed in
both string theory and brane world cosmology, this is not the case. In
string theory, extra dimensions are a theoretical necessity since
superstring theory requires a ten-dimensional spacetime to be consistent
from the quantum point of view, while in brane world cosmology matter and
gauge interactions are localized on a 3-brane, embedded into a higher
dimensional spacetime in which gravity propagates throughout the whole of
spacetime. The most natural extension of general relativity in higher
dimensional spacetimes with the assumption of Einstein -- that the left hand
side of the field equations is the most general symmetric conserved tensor
containing no more than two derivatives of the metric -- is Lovelock theory.
Lovelock \cite{Lov} found the most general symmetric conserved tensor
satisfying this property. The resultant tensor is nonlinear in the Riemann
tensor and differs from the Einstein tensor only if the spacetime has more
than 4 dimensions. Since the Lovelock tensor contains metric derivatives no
higher second order, the quantization of the linearized Lovelock theory is
ghost-free \cite{Zw}. The concepts of action and energy-momentum play
central roles in gravity. However there is no good local notion of energy
for a gravitating system. Quasilocal definitions of energy and conserved
quantities for Einstein gravity \cite{BY,BCM,BoothMann} define a stress
energy tensor on the boundary of some region within the spacetime through
the use of the well-defined gravitational action of Einstein gravity with
the surface term of Gibbons and Hawking \cite{Gib}. Our first aim in this
paper is to generalize the definition of the quasilocal stress energy tensor
for computing the conserved quantities of a solution of third order Lovelock
gravity with zero curvature boundary. The first step is to find the surface
terms for the action of third order Lovelock gravity that make the action
well-defined. These surface terms were introduced by Myers in terms of
differential forms \cite{Myers}. The explicit form of the surface terms for
second order Lovelock gravity has been written in Ref. \cite{Dav}. Here, we
write down the tensorial form of the surface term for the third order
Lovelock gravity, and then obtain the stress energy tensor via the
quasilocal formalism. Of course, as in the case of Einstein gravity, the
action and conserved quantities diverge when the boundary goes to infinity.
We will also introduce a counterterm to deal with these divergences. This is
quite straightforward for the cases we consider in which the boundary is
flat. This is because all curvature invariants are zero except for a
constant, and so the only possible boundary counterterm is one proportional
to the volume of the boundary regardless of the number of dimensions. The
coefficient of this volume counterterm is the same for solutions with flat
or curved boundary. The issue of determination of boundary counterterms with
their coefficients for higher-order Lovelock theories is at this point an
open question. Since the Lovelock Lagrangian appears in the low energy limit
of string theory, there has in recent years been a renewed interest in
Lovelock gravity. In particular, exact static spherically symmetric black
hole solutions of the Gauss-Bonnet gravity (quadratic in the Riemann tensor)
have been found in Ref. \cite{Des}, and of the Maxwell-Gauss-Bonnet and
Born-Infeld-Gauss-Bonnet models in Ref. \cite{Wil1}. The thermodynamics of
the uncharged static spherically black hole solutions has been considered in
\cite{MS}, of solutions with nontrivial topology and asymptotically de
Sitter in \cite{Cai} and of charged solutions in \cite{Wil1,Od1}. Very
recently NUT charged black hole solutions of Gauss-Bonnet gravity and
Gauss-Bonnet-Maxwell gravity were obtained \cite{DM}. All of these known
solutions in Gauss-Bonnet gravity are static. Not long ago one of us
introduced two new classes of rotating solutions of second order Lovelock
gravity and investigated their thermodynamics \cite{Deh1}, and made the
first attempt for finding exact solutions in third order Lovelock gravity
with the quartic terms \cite{Deh2}. Our second aim in this paper is to
obtain rotating asymptotically anti de Sitter (AdS) black holes of third
order Lovelock gravity and investigate their thermodynamics. Apart from
their possible relevance to string theory, it is of general interest to
explore black holes in generalized gravity theories in order to discover
which properties are peculiar to Einstein's gravity, and which are robust
features of all generally covariant theories of gravity. The outline of our
paper is as follows. We give a brief review of the field equations of third
order Lovelock gravity and the counterterm method for calculating conserved
quantities in Sec. \ref{Fiel}. In Sec. \ref{rot} we introduce the $(n+1)$%
-dimensional solutions with a complete set of rotational parameters and
investigate their properties. In Sec. \ref{Therm} we obtain mass, angular
momentum, entropy, temperature, charge, and electric potential of the $(n+1)$%
-dimensional black hole solutions and show that these quantities satisfy the
first law of thermodynamics. We also perform a local stability analysis of
the black holes in the canonical and grand canonical ensembles. We finish
our paper with some concluding remarks.

\section{Field equations\label{Fiel}}

The action of third order Lovelock gravity in the presence of
electromagnetic field may be written as
\begin{equation}
I_{G}=\frac{1}{16\pi }\int_{\mathcal{M}}d^{n+1}x\sqrt{-g}\left( -2\Lambda
+R+\alpha _{2}\mathcal{L}_{2}+\alpha _{3}\mathcal{L}_{3}-F_{\mu \nu }F^{\mu
\nu }\right)  \label{Act1}
\end{equation}
where $\Lambda $ is the cosmological constant, $\alpha _{2}$ and $\alpha
_{3} $ are Gauss-Bonnet and third order Lovelock coefficients, $F_{\mu \nu
}=\partial _{\mu }A_{\nu }-\partial _{\nu }A_{\mu }$ is electromagnetic
tensor field and $A_{\mu }$ is the vector potential. The first term is the
cosmological term, the second term, $R$, is the Einstein term, the third
term is the Gauss-Bonnet Lagrangian given as
\begin{equation}
\mathcal{L}_{2}=R_{\mu \nu \gamma \delta }R^{\mu \nu \gamma \delta }-4R_{\mu
\nu }R^{\mu \nu }+R^{2}  \label{L2}
\end{equation}
and the last term is the third order Lovelock term
\begin{eqnarray}
\mathcal{L}_{3} &=&2R^{\mu \nu \sigma \kappa }R_{\sigma \kappa \rho \tau }R_{%
\phantom{\rho \tau }{\mu \nu }}^{\rho \tau }+8R_{\phantom{\mu \nu}{\sigma
\rho}}^{\mu \nu }R_{\phantom {\sigma \kappa} {\nu \tau}}^{\sigma \kappa }R_{%
\phantom{\rho \tau}{ \mu \kappa}}^{\rho \tau }+24R^{\mu \nu \sigma \kappa
}R_{\sigma \kappa \nu \rho }R_{\phantom{\rho}{\mu}}^{\rho }  \label{L3} \\
&&+3RR^{\mu \nu \sigma \kappa }R_{\sigma \kappa \mu \nu }+24R^{\mu \nu
\sigma \kappa }R_{\sigma \mu }R_{\kappa \nu }+16R^{\mu \nu }R_{\nu \sigma
}R_{\phantom{\sigma}{\mu}}^{\sigma }-12RR^{\mu \nu }R_{\mu \nu }+R^{3}
\notag
\end{eqnarray}
From a geometric point of view the combination of these terms in seven and
eight dimensions is the most general Lagrangian that yields second order
field equations, as in the four-dimensional case for which the
Einstein-Hilbert action is the most general Lagrangian producing second
order field equations, or the five- and six-dimensional cases, for which the
Einstein-Gauss-Bonnet Lagrangian is the most general one fulfilling this
criterion. Since the third Lovelock term in eq. (\ref{Act1}) is an Euler
density in six dimensions and has no contribution to the field equations in
six or less dimensional spacetimes, we therefore consider $(n+1)$%
-dimensional spacetimes with $n\geq 6$. Varying the action with respect to
the metric tensor $g_{\mu \nu }$ and electromagnetic tensor field $F_{\mu
\nu }$ the equations of gravitation and electromagnetic fields are obtained
as:
\begin{eqnarray}
&&G_{\mu \nu }^{(1)}+\Lambda g_{\mu \nu }+\alpha _{2}G_{\mu \nu
}^{(2)}+\alpha _{3}G_{\mu \nu }^{(3)}=T_{\mu \nu }  \label{Geq} \\
&&\nabla _{\nu }F^{\mu \nu }=0  \label{EMeq}
\end{eqnarray}
where $T_{\mu \nu }=2F_{\phantom{\lambda}{\mu}}^{\rho }F_{\rho \nu }-\frac{1%
}{2}F_{\rho \sigma }F^{\rho \sigma }g_{\mu \nu }$ is the energy-momentum
tensor of electromagnetic field, $G_{\mu \nu }^{(1)}$ is the Einstein
tensor, and $G_{\mu \nu }^{(2)}$ and $G_{\mu \nu }^{(3)}$ are the second and
third order Lovelock tensors given as \cite{Hoi}:
\begin{equation}
G_{\mu \nu }^{(2)}=2(R_{\mu \sigma \kappa \tau }R_{\nu }^{\phantom{\nu}%
\sigma \kappa \tau }-2R_{\mu \rho \nu \sigma }R^{\rho \sigma }-2R_{\mu
\sigma }R_{\phantom{\sigma}\nu }^{\sigma }+RR_{\mu \nu })-\frac{1}{2}%
\mathcal{L}_{2}g_{\mu \nu }  \label{Love2}
\end{equation}
\begin{eqnarray}
G_{\mu \nu }^{(3)} &=&-3(4R^{\tau \rho \sigma \kappa }R_{\sigma \kappa
\lambda \rho }R_{\phantom{\lambda }{\nu \tau \mu}}^{\lambda }-8R_{%
\phantom{\tau \rho}{\lambda \sigma}}^{\tau \rho }R_{\phantom{\sigma
\kappa}{\tau \mu}}^{\sigma \kappa }R_{\phantom{\lambda }{\nu \rho \kappa}%
}^{\lambda }+2R_{\nu }^{\phantom{\nu}{\tau \sigma \kappa}}R_{\sigma \kappa
\lambda \rho }R_{\phantom{\lambda \rho}{\tau \mu}}^{\lambda \rho }  \notag \\
&&-R^{\tau \rho \sigma \kappa }R_{\sigma \kappa \tau \rho }R_{\nu \mu }+8R_{%
\phantom{\tau}{\nu \sigma \rho}}^{\tau }R_{\phantom{\sigma \kappa}{\tau \mu}%
}^{\sigma \kappa }R_{\phantom{\rho}\kappa }^{\rho }+8R_{\phantom
{\sigma}{\nu \tau \kappa}}^{\sigma }R_{\phantom {\tau \rho}{\sigma \mu}%
}^{\tau \rho }R_{\phantom{\kappa}{\rho}}^{\kappa }  \notag \\
&&+4R_{\nu }^{\phantom{\nu}{\tau \sigma \kappa}}R_{\sigma \kappa \mu \rho
}R_{\phantom{\rho}{\tau}}^{\rho }-4R_{\nu }^{\phantom{\nu}{\tau \sigma
\kappa }}R_{\sigma \kappa \tau \rho }R_{\phantom{\rho}{\mu}}^{\rho
}+4R^{\tau \rho \sigma \kappa }R_{\sigma \kappa \tau \mu }R_{\nu \rho
}+2RR_{\nu }^{\phantom{\nu}{\kappa \tau \rho}}R_{\tau \rho \kappa \mu }
\notag \\
&&+8R_{\phantom{\tau}{\nu \mu \rho }}^{\tau }R_{\phantom{\rho}{\sigma}%
}^{\rho }R_{\phantom{\sigma}{\tau}}^{\sigma }-8R_{\phantom{\sigma}{\nu \tau
\rho }}^{\sigma }R_{\phantom{\tau}{\sigma}}^{\tau }R_{\mu }^{\rho }-8R_{%
\phantom{\tau }{\sigma \mu}}^{\tau \rho }R_{\phantom{\sigma}{\tau }}^{\sigma
}R_{\nu \rho }-4RR_{\phantom{\tau}{\nu \mu \rho }}^{\tau }R_{\phantom{\rho}%
\tau }^{\rho }  \notag \\
&&+4R^{\tau \rho }R_{\rho \tau }R_{\nu \mu }-8R_{\phantom{\tau}{\nu}}^{\tau
}R_{\tau \rho }R_{\phantom{\rho}{\mu}}^{\rho }+4RR_{\nu \rho }R_{%
\phantom{\rho}{\mu }}^{\rho }-R^{2}R_{\nu \mu })-\frac{1}{2}\mathcal{L}%
_{3}g_{\mu \nu }  \label{Love3}
\end{eqnarray}
The Einstein-Hilbert action (with $\alpha _{2}=\alpha _{3}=0$) does not have
a well-defined variational principle, since one encounters a total
derivative that produces a surface integral involving the derivative of $%
\delta g_{\mu \nu }$ normal to the boundary. These normal derivative terms
do not vanish by themselves, but are canceled by the variation of the
Gibbons-Hawking surface term \cite{Gib}
\begin{equation}
I_{b}^{(1)}=\frac{1}{8\pi }\int_{\delta \mathcal{M}}d^{n}x\sqrt{-\gamma }K
\label{Ib1}
\end{equation}
The main difference between higher derivative gravity and Einstein gravity
is that the surface term that renders the variational principle well-behaved
is much more complicated. However, the surface terms that make the
variational principle well-defined are known for the case of Gauss-Bonnet
gravity\cite{Myers, Dav} to be $I_{b}^{(1)}+I_{b}^{(2)}$, where $I_{b}^{(2)}$
is
\begin{equation}
I_{b}^{(2)}=\frac{1}{8\pi }\int_{\delta \mathcal{M}}d^{n}x\sqrt{-\gamma }%
\left\{ 2\alpha _{2}\left( J-2\widehat{G}_{ab}^{(1)}K^{ab}\right) \right\}
\label{Ib2}
\end{equation}
and where $\gamma _{\mu \nu }$ is induced metric on the boundary, $K$ is
trace of extrinsic curvature of boundary, $\widehat{G}_{ab}^{(1)}$ is the $n$%
-dimensional Einstein tensor of the metric $\gamma _{ab}$ and $J$ is the
trace of
\begin{equation}
J_{ab}=\frac{1}{3}%
(2KK_{ac}K_{b}^{c}+K_{cd}K^{cd}K_{ab}-2K_{ac}K^{cd}K_{db}-K^{2}K_{ab})
\label{Jab}
\end{equation}
For the case of third order Lovelock gravity, the surface term that makes
the variational principle well defined is $%
I_{b}=I_{b}^{(1)}+I_{b}^{(2)}+I_{b}^{(3)}$, where $I_{b}^{(3)}$ is
\begin{eqnarray}
&&I_{b}^{(3)}=\frac{1}{8\pi }\int_{\delta \mathcal{M}}d^{n}x\sqrt{-\gamma }%
\{3\alpha _{3}(P-2\widehat{G}_{ab}^{(2)}K^{ab}-12\widehat{R}_{ab}J^{ab}+2%
\widehat{R}J  \notag \\
&&\hspace{3cm}-4K\widehat{R}_{abcd}K^{ac}K^{bd}-8\widehat{R}%
_{abcd}K^{ac}K_{e}^{b}K^{ed})\}  \label{Ib3}
\end{eqnarray}
In eq. (\ref{Ib3}) $\widehat{G}_{ab}^{(2)}$ is the second order Lovelock
tensor (\ref{Love2}) for the boundary metric $\gamma _{ab}$, and $P$ is the
trace of
\begin{eqnarray}
P_{ab} &=&\frac{1}{5}%
\{[K^{4}-6K^{2}K^{cd}K_{cd}+8KK_{cd}K_{e}^{d}K^{ec}-6K_{cd}K^{de}K_{ef}K^{fc}+3(K_{cd}K^{cd})^{2}]K_{ab}
\notag \\
&&-(4K^{3}-12KK_{ed}K^{ed}+8K_{de}K_{f}^{e}K^{fd})K_{ac}K_{b}^{c}-24KK_{ac}K^{cd}K_{de}K_{b}^{e}
\notag \\
&&+(12K^{2}-12K_{ef}K^{ef})K_{ac}K^{cd}K_{db}+24K_{ac}K^{cd}K_{de}K^{ef}K_{bf}\}
\label{Pab}
\end{eqnarray}
In general $I_{G}+I_{b}^{(1)}+I_{b}^{(2)}+I_{b}^{(3)}$ is divergent when
evaluated on solutions, as is the Hamiltonian and other associated conserved
quantities \cite{BY,BCM,BoothMann}. One way of eliminating these divergences
is through the use of background subtraction \cite{BY}, in which the
boundary surface is embedded in another (background) spacetime, and all
quasilocal quantities are computed with respect to this background,
incorporated into the theory by adding to the action the extrinsic curvature
of the embedded surface. Such a procedure causes the resulting physical
quantities to depend on the choice of reference background; furthermore, it
is not possible in general to embed the boundary surface into a background
spacetime. For asymptotically AdS solutions, one can instead deal with these
divergences via the counterterm method inspired by AdS/CFT correspondence
\cite{Mal}. This conjecture, which relates the low energy limit of string
theory in asymptotically anti de-Sitter spacetime and the quantum field
theory on its boundary, has attracted a great deal of attention in recent
years. The equivalence between the two formulations means that, at least in
principle, one can obtain complete information on one side of the duality by
performing computation on the other side. A dictionary translating between
different quantities in the bulk gravity theory and their counterparts on
the boundary has emerged, including the partition functions of both
theories. In the present context this conjecture furnishes a means for
calculating the action and conserved quantities intrinsically without
reliance on any reference spacetime \cite{Sken,BK,Od2} by adding additional
terms on the boundary that are curvature invariants of the induced metric.
Although there may exist a very large number of possible invariants one
could add in a given dimension, only a finite number of them are
nonvanishing as the boundary is taken to infinity. Its many applications
include computations of conserved quantities for black holes with rotation,
NUT charge, various topologies, rotating black strings with zero curvature
horizons and rotating higher genus black branes \cite{Deh3}. Although the
counterterm method applies for the case of a specially infinite boundary, it
was also employed for the computation of the conserved and thermodynamic
quantities in the case of a finite boundary \cite{DM2}. Extensions to de
Sitter spacetime and asymptotically flat spacetimes have also been proposed
\cite{dS}. All of the work mentioned in the previous paragraph was limited
to Einstein gravity. Here we apply the counterterm method to the case of the
solutions of the field equations of third order Lovelock gravity. At any
given dimension there are only finitely many counterterms that one can write
down that do not vanish at infinity. This does not depend upon what the bulk
theory is -- i.e. whether or not it is Einstein, Gauss-Bonnet, 3rd order
Lovelock, etc. Indeed, for asymptotically (A)dS solutions, the boundary
counterterms that cancel divergences in Einstein Gravity should also cancel
divergences in 2nd and 3rd order Lovelock gravity. The coefficients will be
different and depends on $\Lambda $\ and Lovelock coefficients as we will
see this for the volume term in the flat boundary case below. Of course
these coefficients should reduce to those in Einstein gravity as one may
expect. Unfortunately we do not have a rotating solution to either
Gauss-Bonnet or 3rd-order Lovelock gravity that does not have a flat
boundary at infinity. Consequently we restrict our considerations to
counterterms for the flat-boundary case, i.e. $\widehat{R}_{abcd}(\gamma )=0$%
, for which there exists only one boundary counterterm
\begin{equation}
I_{ct}=\frac{1}{8\pi }\int_{\delta \mathcal{M}}d^{n}x\sqrt{-\gamma }\frac{n-1%
}{L},  \label{Ict}
\end{equation}
where $L$ is a scale length factor that depends on $l$, $\alpha _{2}$ and $%
\alpha _{3}$, that must reduce to $l$ as $\alpha _{2}$ and $\alpha _{3}$ go
to zero. Having the total finite action $%
I=I_{G}+I_{b}^{(1)}+I_{b}^{(2)}+I_{b}^{(3)}$, one can use the quasilocal
definition \cite{BY,BCM} to construct a divergence free stress-energy
tensor. For the case of manifolds with zero curvature boundary the finite
stress energy tensor is
\begin{eqnarray}
T^{ab} &=&\frac{1}{8\pi }\{(K^{ab}-K\gamma ^{ab})+2\alpha
_{2}(3J^{ab}-J\gamma ^{ab})  \notag \\
&&\ +3\alpha _{3}(5P^{ab}-P\gamma ^{ab})+\frac{n-1}{L}\gamma ^{ab}\ \}.
\label{Stres}
\end{eqnarray}
The first three terms in eq. (\ref{Stres}) result from the variation of the
surface action (\ref{Ib1})-(\ref{Pab}) with respect to $\gamma ^{ab}$, and
the last term is the counterterm that is the variation of $I_{ct}$ with
respect to $\gamma ^{ab}$. To compute the conserved charges of the
spacetime, we choose a spacelike surface $\mathcal{B}$ in $\partial \mathcal{%
M}$ with metric $\sigma _{ij}$, and write the boundary metric in ADM form:
\begin{equation}
\gamma _{ab}dx^{a}dx^{a}=-N^{2}dt^{2}+\sigma _{ij}\left( d\varphi
^{i}+V^{i}dt\right) \left( d\varphi ^{j}+V^{j}dt\right) ,
\end{equation}
where the coordinates $\varphi ^{i}$ are the angular variables
parameterizing the hypersurface of constant $r$ around the origin, and $N$
and $V^{i}$ are the lapse and shift functions respectively. When there is a
Killing vector field $\mathcal{\xi }$ on the boundary, then the quasilocal
conserved quantities associated with the stress tensors of eq. (\ref{Stres})
can be written as
\begin{equation}
\mathcal{Q}(\mathcal{\xi )}=\int_{\mathcal{B}}d^{n-1}\varphi \sqrt{\sigma }%
T_{ab}n^{a}\mathcal{\xi }^{b},  \label{charge}
\end{equation}
where $\sigma $ is the determinant of the metric $\sigma _{ij}$, and $n^{a}$
is the timelike unit normal vector to the boundary $B$\textbf{.} For
boundaries with timelike ($\xi =\partial /\partial t$) and rotational ($%
\varsigma =\partial /\partial \varphi $) Killing vector fields, one obtains
the quasilocal mass and angular momentum
\begin{eqnarray}
M &=&\int_{\mathcal{B}}d^{n-1}\varphi \sqrt{\sigma }T_{ab}n^{a}\xi ^{b},
\label{Mtot} \\
J &=&\int_{\mathcal{B}}d^{n-1}\varphi \sqrt{\sigma }T_{ab}n^{a}\varsigma
^{b},  \label{Jtot}
\end{eqnarray}
provided the surface $\mathcal{B}$ contains the orbits of $\varsigma $.
These quantities are, respectively, the conserved mass and angular momentum
of the system enclosed by the boundary $\mathcal{B}$. Note that they will
both depend on the location of the boundary $\mathcal{B}$ in the spacetime,
although each is independent of the particular choice of foliation $\mathcal{%
B}$ within the surface $\partial \mathcal{M}$.

\section{$(n+1)$-dimensional Rotating Solutions\label{rot}}

As stated before, the third order Lovelock term in eq. (\ref{Act1}) is an
Euler density in six dimensions and has no contribution to the field
equations in spacetimes of dimension six or less. Taking $n\geq 6$, we
obtain the $(n+1)$-dimensional solutions of third order Lovelock gravity
with nonvanishing electromagnetic field with $k$ rotation parameters and
investigate their properties. The rotation group in $n+1$ dimensions is $%
SO(n)$ and therefore the number of independent rotation parameters is $%
[(n+1)/2]$, where $[x]$ is the integer part of $x$. The metric of an $(n+1)$%
-dimensional asymptotically AdS rotating solution with $k\leq \lbrack
(n+1)/2]$ rotation parameters whose constant $(t,r)$ hypersurface has zero
curvature may be written as \cite{Awad}
\begin{eqnarray}
ds^{2} &=&-f(r)\left( \Xi dt-{{\sum_{i=1}^{k}}}a_{i}d\phi _{i}\right) ^{2}+%
\frac{r^{2}}{l^{4}}{{\sum_{i=1}^{k}}}\left( a_{i}dt-\Xi l^{2}d\phi
_{i}\right) ^{2}  \notag \\
&&\ +\frac{dr^{2}}{f(r)}-\frac{r^{2}}{l^{2}}{\sum_{i<j}^{k}}(a_{i}d\phi
_{j}-a_{j}d\phi _{i})^{2}+r^{2}dX^{2},  \label{met2}
\end{eqnarray}
where $\Xi =\sqrt{1+\sum_{i}^{k}a_{i}^{2}/l^{2}}$, the angular coordinates
are in the range $0\leq \phi _{i}<2\pi $ and $dX^{2}$ is the Euclidean
metric on the $\left( n-k-1\right) $-dimensional submanifold with volume $%
\Sigma _{n-k-1}$. Using eq. (\ref{EMeq}), one can show that the vector
potential can be written as
\begin{equation}
A_{\mu }=\frac{q}{(n-2)r^{n-2}}(\Xi \delta _{\mu }^{0}-a_{i}\delta _{\mu
}^{i}),\hspace{0.5cm}(\mathtt{no\ sum\ on\ i}).
\end{equation}
where $q$ is an arbitrary real constant which is related to the charge of
the solution. To find the function $f(r)$, one may use any components of eq.
(\ref{Geq}). The simplest equation is the $rr$ component of these equations
which can be written as
\begin{eqnarray}
&&\left[ 180(_{\phantom{n}{5}}^{n-1})\alpha _{3}rf^{2}-6(_{\phantom{n}{3}%
}^{n-1})\alpha _{2}fr^{3}+\frac{n-1}{2}r^{5})\right] f^{\prime }+\Lambda
r^{6}  \notag \\
&&+360(_{\phantom{n}{6}}^{n-1})\alpha _{3}f^{3}-12(_{\phantom{n}{4}%
}^{n-1})\alpha _{2}r^{2}f^{2}+(_{\phantom{n}{2}}^{n-1})r^{4}f=-q^{2}r^{8-2n}
\label{Eqf}
\end{eqnarray}
where prime denotes the derivative with respect to $r$. We present the full
solution to this equation in the appendix. Here, for simplicity, we consider
the solutions of eq. (\ref{Eqf}) for a restricted version of the $\alpha
_{i} $'s given as
\begin{equation}
\alpha _{3}=\frac{\alpha ^{2}}{72(_{\phantom{2}{4}}^{n-2})},\hspace{1cm}%
\alpha _{2}=\frac{\alpha }{(n-2)(n-3)}  \label{con}
\end{equation}
Equation (\ref{Eqf}) with condition (\ref{con}) has one real and two complex
solutions that are the complex conjugate of each other. The real solution of
eq. (\ref{Eqf}) with condition (\ref{con}) is
\begin{equation}
f(r)={\frac{{r}^{2}}{\alpha }}\left\{ 1-\left( {1+\frac{6\Lambda \alpha }{%
n(n-1)}+\frac{3\alpha m}{r^{n}}-\frac{6\alpha {q}^{2}}{\left( n-1\right)
(n-2)r^{2n-2}}}\right) ^{1/3}\right\}  \label{Frs}
\end{equation}
where $m$ is the mass parameter. Although the other components of the field
equation (\ref{Geq}) are more complicated, one can check that the metric (%
\ref{met2}) satisfies all the components eq. (\ref{Geq}) provided above $%
f(r) $ is given by (\ref{Frs}). Unlike the solution in Gauss-Bonnet gravity,
which has two branches, the solution (\ref{Frs}) has only one branch.
Indeed, eq. (\ref{Eqf}) with the above $\alpha $'s has the real solution (%
\ref{Frs}) and two complex solutions that are complex conjugates of each
other. This feature is the same as the asymptotically AdS solution of Ref.
\cite{ATZ}, which has a unique anti de Sitter vacuum. These solutions are
asymptotically AdS or dS for negative or positive values of $\Lambda $
respectively. In this paper we are interested in the case of asymptotically
AdS solutions, and therefore we put $\Lambda =-n(n-1)/2l^{2}$. One can show
that the Kretschmann scalar $R_{\mu \nu \lambda \kappa }R^{\mu \nu \lambda
\kappa }$ diverges at $r=0$, and therefore there is a curvature singularity
located at $r=0$. Seeking possible black hole solutions, we turn to looking
for the existence of horizons. As in the case of rotating black hole
solutions of Einstein gravity, the above metric given by eqs. (\ref{met2})
and (\ref{Fr}) has both Killing and event horizons. The Killing horizon is a
null surface whose null generators are tangent to a Killing field. The proof
that a stationary black hole event horizon must be a Killing horizon in the
four-dimensional Einstein gravity \cite{Haw1} cannot obviously be
generalized to higher order gravity. However the result is true for all
known static solutions. Although our solution is not static, the Killing
vector
\begin{equation}
\chi =\partial _{t}+{{{\sum_{i=1}^{k}}}}\Omega _{i}\partial _{\phi _{i}},
\label{Kil}
\end{equation}
is the null generator of the event horizon, where $k$ denotes the number of
rotation parameters. The event horizon is defined by the solution of $%
g^{rr}=f(r)=0$. For the case of uncharged solutions, there exists only one
event horizon located at
\begin{equation}
r_{+}=(ml^{2})^{1/n}  \label{rhq0}
\end{equation}
This feature is different from the case of uncharged spherically symmetric
solutions of third order Lovelock gravity with curved horizon \cite{Deh2},
for which one may have uncharged black holes with two horizons, extreme
ones, or a naked singularity. As we demonstrate in the appendix, the general
uncharged solution does not have two horizons\textbf{.} The charged solution
presents a black hole solution with two inner and outer horizons, provided
the mass parameter $m$ is greater than, $m_{\mathrm{ext}}$, an extreme black
hole for $m=m_{\mathrm{ext}}$ and a naked singularity otherwise, where $m_{%
\mathrm{ext}}$ is
\begin{equation}
m_{\mathrm{ext}}=\frac{2(n-1)}{(n-2)l^{2}}\left( \frac{2q^{2}l^{2}}{n(n-1)}%
\right) ^{n/2(n-1)}  \label{mext}
\end{equation}
\textbf{\ }The general charged solution also has only two horizons, as shown
in the appendix.

\section{Thermodynamics of black holes \label{Therm}}

One can obtain the temperature and angular momentum of the event horizon by
analytic continuation of the metric. Setting $t\rightarrow i\tau $ and $%
a_{i}\rightarrow ia_{i}$ yields the Euclidean section of (\ref{met2}), whose
regularity at $r=r_{+}$ requires that we should identify $\tau \sim \tau
+\beta _{+}$ and $\phi _{i}\sim \phi _{i}+\beta _{+}\Omega _{i}$, where $%
\beta _{+}$ and $\Omega _{i}$'s are the inverse Hawking temperature and the
angular velocities of the outer event horizon. One obtains
\begin{eqnarray}
\beta _{+}^{-1} &=&T_{+}=\frac{f^{\prime }(r_{+})}{4\pi \Xi }=\frac{%
n(n-1)-2l^{2}q^{2}r_{+}^{-2(n-1)}}{4\pi (n-1)\Xi l^{2}}r_{+},  \label{bet1}
\\
\Omega _{i} &=&\frac{a_{i}}{\Xi l^{2}}.  \label{Om1}
\end{eqnarray}
Next, we calculate the electric charge of the solutions. To determine the
electric field we should consider the projections of the electromagnetic
field tensor on special hypersurfaces. The normal to such hypersurfaces is
\begin{equation}
u^{0}=\frac{1}{N},\hspace{0.5cm}u^{r}=0,u^{i}=\frac{V^{i}}{N},
\end{equation}
and the electric field is $E^{\mu }=g^{\mu \rho }F_{\rho \nu }u^{\nu }$.
Then the electric charge per unit volume $V_{n-1}$ can be found by
calculating the flux of the electric field at infinity, yielding
\begin{equation}
Q=\frac{\Xi }{4\pi }q  \label{Ch}
\end{equation}
The electric potential $\Phi $, measured at infinity with respect to the
horizon, is defined by \cite{Gub}
\begin{equation}
\Phi =A_{\mu }\chi ^{\mu }\left| _{r\rightarrow \infty }-A_{\mu }\chi ^{\mu
}\right| _{r=r_{+}},  \label{Pot1}
\end{equation}
where $\chi $ is the null generator of the horizon given by eq. (\ref{Kil}).
We find
\begin{equation}
\Phi =\frac{q}{(n-2)\Xi r_{+}^{n-2}}  \label{Pot2}
\end{equation}
Conserved quantities associated with the spacetime described by (\ref{met2})
can be obtained via the counterterm method. Using eqs. (\ref{Mtot}) and (\ref
{Jtot}), the mass and angular momentum will be finite provided
\begin{eqnarray*}
L &=&\frac{15l^{2}\sqrt{\alpha (1-\lambda )}}{5l^{2}+9\alpha -l^{2}\lambda
^{2}-4l^{2}\lambda } \\
\lambda  &=&\left( 1-\frac{3\alpha }{l^{2}}\right) ^{1/3}
\end{eqnarray*}
where we note that $L$ reduces to $l$ as $\alpha $ goes to zero. The mass
and angular momentum per unit volume $V_{n-1}$ can then be obtained through
the use of eqs. (\ref{Mtot}) and (\ref{Jtot}). We find
\begin{eqnarray}
M &=&\frac{1}{16\pi }[n\Xi ^{2}-1]m,  \label{Mtot1} \\
J_{i} &=&\frac{1}{16\pi }n\Xi ma_{i}  \label{Jtot1}
\end{eqnarray}
Black hole entropy typically satisfies the so-called area law, which states
that the entropy of a black hole equals one-quarter of the area of its
horizon \cite{Beck}. This near-universal law applies to all kinds of black
holes and black strings in Einstein gravity \cite{Haw2}. However in higher
derivative gravity the area law is not satisfied in general \cite{fails}. It
is known in Lovelock gravity that \cite{Myers2,Ross}
\begin{equation}
S=\frac{1}{4}\sum_{k=1}^{[(d-1)/2]}k\alpha _{k}\int d^{n-1}x\sqrt{\tilde{g}}%
\tilde{\mathcal{L}}_{k-1}  \label{Enta}
\end{equation}
where the integration is done on the $(n-1)$-dimensional spacelike
hypersurface of Killing horizon, $\tilde{g}_{\mu \nu }$ is the induced
metric on it, $\tilde{g}$ is the determinant of $\tilde{g}_{\mu \nu }$ and $%
\tilde{\mathcal{L}}_{k}$ is the $k$th order Lovelock Lagrangian of $\tilde{g}%
_{\mu \nu }$. For the topological class of black holes we are considering,
the horizon curvature is zero. Consequently the area law holds. Denoting the
volume of the hypersurface boundary at constant $t$ and $r$ by $%
V_{n-1}=(2\pi )^{k}\Sigma _{n-k-1}$, we obtain
\begin{equation}
S=\frac{\Xi }{4}r_{+}^{n-1}  \label{Ent1}
\end{equation}
for the entropy per unit volume $V_{n-1}$. \ The entropy can also be
obtained through the use of Gibbs-Duhem relation
\begin{equation}
S=\beta (\mathcal{M}-\Gamma _{i}\mathcal{C}_{i})-I  \label{Gibs}
\end{equation}
where $I$ is the finite total action evaluated on the classical solution,
and $\mathcal{C}_{i}$ and $\Gamma _{i}$ are the conserved charges and their
associate chemical potentials respectively. For simplicity, we consider the
uncharged solutions, for which $\mathcal{C}_{i}=J_{i}$ and $\Gamma
_{i}=\Omega _{i}$. Using eqs. (\ref{Act1}), (\ref{Ib1})-(\ref{Ib3}) and (\ref
{Ict}), the finite total action per unit volume $V_{n-1}$ can be calculated
as
\begin{equation}
I=-\frac{\beta _{+}}{16\pi l^{2}}r_{+}^{n}  \label{Acttot}
\end{equation}
Now using Gibbs-Duhem relation (\ref{Gibs}) and eqs. (\ref{Mtot1}), (\ref
{Jtot1}) and \ref{Acttot}) one may confirm that the entropy per unit volume
obeys the area law of eq. (\ref{Ent1}).

\subsection{Energy as a function of entropy, angular momenta and charge}

We first obtain the mass as a function of the extensive quantities $S$, $%
\mathbf{J}$ and $Q$. Using the expression for the entropy, the mass, the
angular momenta, and the charge given in eqs. (\ref{bet1}), (\ref{Ch}), (\ref
{Mtot1}) and (\ref{Jtot1}), and the fact that $f(r_{+})=0$, one can obtain a
Smarr-type formula as
\begin{equation}
M(S,\mathbf{J},Q)=\frac{\left( nZ-1)\right) J}{nl\sqrt{Z(Z-1)}},
\label{Smar}
\end{equation}
where $J^{2}=\left| \mathbf{J}\right| ^{2}=\sum_{i}^{k}J_{i}^{2}$ and $Z=\Xi
^{2}$ is the positive real root of the following equation
\begin{equation}
\left( Z-1\right) ^{(n-1)}-\frac{Z}{16S^{2}}\left\{ \frac{4\pi (n-1)(n-2)lSJ%
}{n[(n-1)(n-2)S^{2}+2\pi ^{2}Q^{2}l^{2}]}\right\} ^{(2n-2)}=0.  \label{ZSmar}
\end{equation}
One may then regard the parameters $S$, $\mathbf{J}$ and $Q$ as a complete
set of extensive parameters for the mass $M(S,\mathbf{J},Q)$ and define the
intensive parameters conjugate to $S$, $J_{i}$ and $Q$. These quantities are
the temperature, the angular velocities, and the electric potential
\begin{equation}
T=\left( \frac{\partial M}{\partial S}\right) _{Q,\mathbf{J}},\ \ \Omega
_{i}=\left( \frac{\partial M}{\partial J_{i}}\right) _{S,Q},\ \ \Phi =\left(
\frac{\partial M}{\partial Q}\right) _{S,\mathbf{J}}.  \label{DSmar}
\end{equation}
It is a matter of straightforward calculation to show that the intensive
quantities calculated by eqs. (\ref{Smar})-(\ref{DSmar}) are the same as
those found earlier in this section. Thus, the thermodynamic quantities
calculated in this section satisfy the first law of thermodynamics,
\begin{equation}
dM=TdS+{{{\sum_{i=1}^{k}}}}\Omega _{i}dJ_{i}+\Phi dQ.  \label{Flth}
\end{equation}

\subsection{Stability in the canonical and the grand-canonical ensemble}

The stability of a thermodynamic system with respect to the small variations
of the thermodynamic coordinates, is usually performed by analyzing the
behavior of the entropy $S(M,Q,\mathbf{J})$ near equilibrium. The local
stability in any ensemble requires that $S(M,Q,\mathbf{J})$ be a concave
function of its extensive variables or that its Legendre transformation is a
convex function of the intensive variables\textbf{.} The stability can also
be studied by the behavior of the energy $M(S,Q,\mathbf{J})$ which should be
a convex function of its extensive variable. Thus, the local stability can
in principle be carried out by finding the determinant of the Hessian matrix
of $M(S,Q,\mathbf{J})$ with respect to its extensive variables $X_{i}$, $%
\mathbf{H}_{X_{i}X_{j}}^{M}=[\partial ^{2}M/\partial X_{i}\partial X_{j}]$
\cite{Gub}. In our case the entropy $S$ is a function of the mass, the
angular momenta, and the charge. The number of thermodynamic variables
depends on the ensemble that is used. In the canonical ensemble, the charge
and the angular momenta are fixed parameters, and therefore the positivity
of the heat capacity $C_{\mathbf{J},Q}=T(\partial S/\partial T)_{\mathbf{J}%
,Q}$ is sufficient to ensure local stability. The heat capacity $C_{Q,%
\mathbf{J}}$ at constant charge and angular momenta is
\begin{equation}
C_{Q,\mathbf{J}}=\frac{\Xi r_{+}^{(n-1)}}{4\Upsilon }[(n-2)\Xi ^{2}+1][%
(n-1)(n-2)r_{+}^{2(n-1)}+2q^{2}l^{2}]\{n(n-1)r_{+}^{2(n-1)}-2q^{2}l^{2}\}
\label{Cap}
\end{equation}
where $\Upsilon $ is
\begin{eqnarray}
\Upsilon &=&4q^{4}l^{4}[(3n-6)\Xi
^{2}-n+3)]-4(n-1)q^{2}l^{2}r_{+}^{(2n-2)}[(3n-6)\Xi ^{2}-n^{2}+3]  \notag \\
&&+n(n-1)^{2}(n-2)r_{+}^{(4n-4)}[(n+2)\Xi ^{2}-(n+1)]  \label{Cap2}
\end{eqnarray}
\begin{figure}[tbp]
\epsfxsize=10cm \centerline{\epsffile{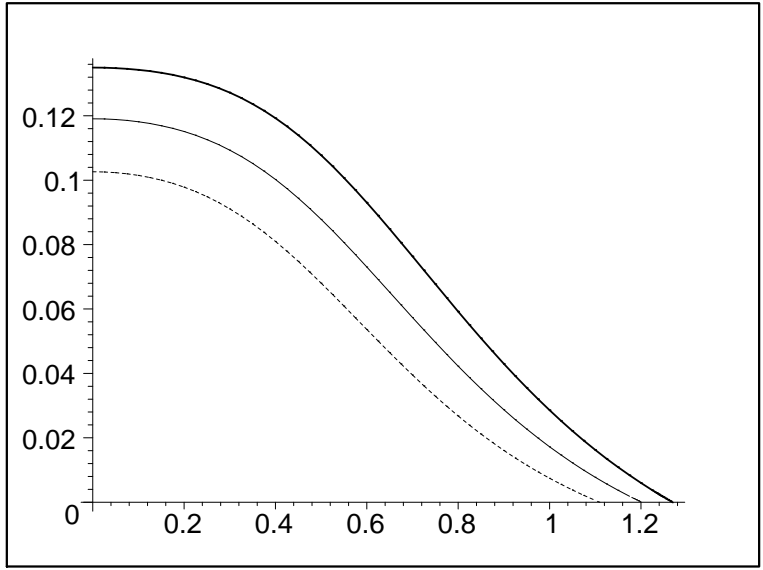}} \caption{$C_{{\bf
J},Q}$ versus $q$ for $l=1$, $r_+=0.8$, $n=6$ (bold-line), $n=7$
(solid-line), and $n=8$ (dotted-line).} \label{Figure1}
\end{figure}
Figure \ref{Figure1} shows the behavior of the heat capacity as a
function of the charge parameter. We see that $C_{Q,\mathbf{J}}$
is positive in various dimensions and goes to zero as $q$
approaches its extreme value. Thus, the $(n+1)$-dimensional
asymptotically AdS charged rotating black brane is locally stable
in the canonical ensemble. In the grand-canonical ensemble, after
some algebraic manipulation, we obtain
\begin{equation}
\left| \mathbf{H}_{S,Q,\mathbf{J}}^{M}\right| =\frac{128\pi }{n[(n-2)\Xi
^{2}+1]l^{2}\Xi ^{6}r_{+}^{3n-4}}\left[ \frac{%
n(n-1)r_{+}^{2(n-1)}+2(2n-3)l^{2}q^{2}}{(n-1)(n-2)r_{+}^{2(n-1)}+2l^{2}q^{2}}%
\right] .  \label{dHes}
\end{equation}
As one can see from eq. (\ref{dHes}), $\left| \mathbf{H}_{S,Q,\mathbf{J}%
}^{M}\right| $ is positive over all phase space. Hence the $(n+1)$%
-dimensional asymptotically AdS charged rotating black brane in third order
Lovelock gravity is locally stable in the grand-canonical ensemble.

\section{Concluding Remarks}

In this paper, first, we introduced the surface terms for the third order
Lovelock gravity which make the action well-defined. This is achieved by
generalizing the Gibbons-Hawking surface term for Einstein gravity or
generalizing the surface terms of Gauss-Bonnet gravity. We generalized the
stress energy momentum tensor of Brown and York in Einstein gravity for the
third order Lovelock gravity for spacetimes with zero curvature at the
boundary. As in the case of Einstein gravity, $I_{G}$, $I_{b}^{(1)}$, $%
I_{b}^{(2)}$ and $I_{b}^{(3)}$ of eqs. (\ref{Act1}), (\ref{Ib1}), (\ref{Ib2}%
) and (\ref{Ib3}) are divergent when evaluated on the solutions, as is the
Hamiltonian and other associated conserved quantities. We also introduced a
counterterm dependent only on the boundary volume, which removed the
divergences of the action and conserved quantities of this solution of third
order Lovelock gravity. We also found a new class of rotating solutions,
whose hypersurfaces of constant $t$ and $r$\ have zero curvature, in third
order Lovelock gravity in the presence of cosmological constant and
electromagnetic field. These solutions are asymptotically AdS or dS for $%
\Lambda <0$ or $\Lambda >0$ respectively. We obtained solutions for special
values of $\alpha _{2}$ and $\alpha _{3}$ given in eq. (\ref{con}) with
negative cosmological constant. In the absence of an electromagnetic field,
these solutions present black branes with one event horizon. The charged
solutions may be interpreted as black brane solutions with two inner and
outer event horizons for $m>m_{\mathrm{ext}}$, extreme black holes for $m=m_{%
\mathrm{ext}}$ or naked singularity for $m<m_{\mathrm{ext}}$, where $m_{%
\mathrm{ext}}$ is given in eq. (\ref{mext}). We found that the Killing
vectors are the null generators of the event horizon, and therefore the
event horizon is a Killing horizon for the stationary solution of the third
order Lovelock gravity explored in this paper. We computed physical
properties of the brane such as the temperature, the angular velocity, the
entropy, the electric charge and potential. Finally, we obtained a
Smarr-type formula for the mass of the black brane solution as a function of
\ the entropy, the charge and the angular momenta of the black brane and
investigated the first law of thermodynamics. We found that the conserved
and thermodynamics quantities satisfy the first law of thermodynamics. We
also studied the phase behavior of the $(n+1)$-dimensional charged rotating
black branes in third order Lovelock gravity and showed that there is no
Hawking-Page phase transition in spite of the charge and angular momenta of
the branes. Indeed, we calculated the heat capacity and the determinant of
the Hessian matrix of the mass with respect to $S$, $\mathbf{J}$ and $Q$ of
the black branes and found that they are positive for all the phase space,
which means that the brane is stable for all the allowed values of the
metric parameters discussed in Sec. \ref{Therm}. This phase behavior is
commensurate with the fact that there is no Hawking-Page transition for a
black object whose horizon is diffeomorphic to $\mathbb{R}^{p}$ and
therefore the system is always in the high temperature phase \cite{Wit2}.

{\Large Acknowledgments} This work has been supported in part by Research
Institute for Astrophysics and Astronomy of Maragha, Iran and also by the
Natural Sciences and Engineering Research Council of Canada. R. Mann is
grateful to the Kavili Institute for Theoretical Physics where part of this
work was carried out.

\section{Appendix}

The general solution in $n+1$ dimensions is
\begin{equation}
f(r)=\frac{b_{2}r^{2}}{b_{3}\alpha }\left\{ 1-\left( \sqrt{\gamma +k^{2}(r)}%
+k(r)\right) ^{1/3}+b_{2}^{2}\left( \sqrt{\gamma +k^{2}(r)}-k(r)\right)
^{1/3}\right\}  \label{Fr}
\end{equation}
where
\begin{equation}
\alpha _{2}=b_{2}\frac{\alpha }{12}\quad \alpha _{3}=b_{3}\frac{\alpha ^{2}}{%
72}\qquad \gamma =\left( \frac{b_{3}-b_{2}^{2}}{b_{2}^{2}}\right) ^{3}\qquad
\lambda =\frac{1}{2}+\frac{3}{2}\gamma ^{1/3}+\frac{3\alpha \Lambda b_{3}^{2}%
}{n(n-1)}\
\end{equation}
and
\begin{equation}
k(r)=\lambda +3\alpha b_{3}^{2}\left( \frac{m}{2r^{n}}-\frac{q^{2}}{%
(n-1)(n-2)r^{2(n-1)}}\right)  \label{Kr}
\end{equation}
The above $f(r)$ reduces to the solution (\ref{Frs}) when $b_{3}=b_{2}^{2}=1$%
.. At large $r$ the function $k$ approaches a constant, and the spacetime is
asymptotically de Sitter or anti de Sitter depending on the overall sign of
\begin{equation*}
\Lambda _{\mathrm{eff}}=\frac{b_{2}}{\alpha b_{3}}\left\{ 1-\left( \lambda +%
\sqrt{\gamma +\lambda ^{2}}\right) ^{1/3}+b_{2}^{2}{\left( -\lambda +\sqrt{%
\gamma +\lambda ^{2}}\right) ^{1/3}}\right\}
\end{equation*}
where we have assumed $\alpha >0$ without loss of generality. First, we
investigate the conditions of the reality of $f(r)$. In order to have real $%
f(r)$ the expression $\gamma +k^{2}(r)$ should be positive. This occurs for $%
\gamma >0$, but for negative $\gamma $, this holds if $k^{2}(r)>\left|
\gamma \right| $. The analysis of this case proceeds as follows. 1) If $%
\lambda >0$, then $k(r)$ is zero some where and the condition $%
k^{2}(r)>\left| \gamma \right| $\ violated, and therefore the function $f(r)$
is complex near the root(s) of $k(r)=0$. 2) For $\lambda <0$ the condition $%
k^{2}(r)>\left| \gamma \right| $ holds provided
\begin{equation*}
m\leq \frac{4(n-1)}{(n-2)}\left( \frac{\left| \lambda \right| -\sqrt{\left|
\gamma \right| }}{3\alpha b_{3}^{2}}\right) ^{(n-2)/2(n-1)}\left( \frac{q^{2}%
}{n(n-1)}\right) ^{n/2(n-1)}
\end{equation*}
Note that in this case $\left| \lambda \right| $ should be larger than $%
\sqrt{\left| \gamma \right| }$. Seeking possible black hole solutions, we
turn to looking for the existence of horizons. The roots of the metric
function $f(r)$ are located at
\begin{equation}
k(r)=k_{0}\equiv \frac{\left( 1+\sqrt{1+4b_{2}^{2}\gamma ^{3}}\right)
^{6}-64\gamma }{16\left( 1+\sqrt{1+4b_{2}^{2}\gamma ^{3}}\right) ^{3}}
\label{Rh}
\end{equation}
which reduces to the value of $1/2$ when $\gamma =0$ as in eq. (\ref{con}).
Note that $k_{0}$\ is real if $\gamma ^{3}>-1/\left( 4b_{2}^{2}\right) $.
Thus, there is no black hole solution for $\gamma ^{3}<-1/\left(
4b_{2}^{2}\right) $\textbf{. }For the case of uncharged solutions, there
exists only one event horizon located at
\begin{equation}
r_{+}=\left( \frac{m}{2\sigma }\right) ^{1/n},\qquad \sigma =\frac{%
k_{0}-\lambda }{3\alpha b_{3}^{2}}  \label{rhq0gen}
\end{equation}
provided $\sigma $ is positive. If $\sigma $ is negative or zero then there
will be no roots (corresponding to a naked singularity). Note that $\sigma
=1/2l^{2}$ when $\gamma =0$ and therefore the above equation reduces to eq. (%
\ref{rhq0}). Again this feature is different from the case of uncharged
spherically symmetric solutions of third order Lovelock gravity with curved
horizon \cite{Deh2}, for which one may have uncharged black holes with two
horizons, extreme ones, or a naked singularity. The charged solution
presents a black hole solution with two inner and outer horizons, provided
the mass parameter $m$ is greater than $m_{\mathrm{ext}}$, an extreme black
hole for $m=m_{\mathrm{ext}}$ and a naked singularity otherwise, where $m_{%
\mathrm{ext}}$ is
\begin{equation}
m_{\mathrm{ext}}=\frac{4(n-1)\sigma }{(n-2)}\left( \frac{q^{2}}{n(n-1)\sigma
}\right) ^{n/2(n-1)}  \label{mextgen}
\end{equation}
Again, one may note that for $\gamma =0$ ($\sigma =1/2l^{2}$) eq. (\ref
{mextgen}) reduce to $m_{\mathrm{ext}}$ given in eq. (\ref{mext}).


\begin{thebibliography}{99}
\bibitem{Lov}  D. Lovelock, J. Math. Phys. \textbf{12}, 498 (1971); N.
Deruelle and L. Farina-Busto, Phys. Rev. D \textbf{41}, 3696 (1990); G. A.
MenaMarugan, \emph{ibid}. \textbf{46}, 4320 (1992); 4340 (1992).

\bibitem{Zw}  B. Zwiebach, Phys. Lett. \textbf{B156}, 315 (1985); B. Zumino,
Phys. Rep. \textbf{137}, 109 (1986).

\bibitem{BY}  J.D. Brown and J.W. York, Phys. Rev. \textbf{D47}, 1407 (1993).

\bibitem{BCM}  J. D. Brown, J. Creighton and R. B. Mann, Phys. Rev. \textbf{%
D50}, 6394 (1994).

\bibitem{BoothMann}  I. S. Booth and R. B. Mann, Phys. Rev. D \textbf{59},
064021 (1999).

\bibitem{Gib}  G. W. Gibbons and S. W. Hawking, Phys. Rev. D \textbf{15},
2752 (1977).

\bibitem{Myers}  R. C. Myers, Phys. Rev. D \textbf{36}, 392 (1987).

\bibitem{Dav}  S. C. Davis, Phys. Rev. D \textbf{67}, 024030 (2003).

\bibitem{Des}  D. G. Boulware and S. Deser, Phys. Rev. Lett., \textbf{55},
2656 (1985); J. T. Wheeler, Nucl. Phys. \textbf{B268}, 737 (1986).

\bibitem{Wil1}  D. L. Wiltshire, Phys. Lett. B \textbf{169}, 36 (1986);
Phys. Rev. D \textbf{38}, 2445 (1988).

\bibitem{MS}  R. C. Myers and J. Z. Simon, Phys. Rev. D \textbf{38}, 2434
(1988); R. C. Myers, Nucl. Phys. \textbf{B289}, 701 (1987).

\bibitem{Cai}  R. G. Cai, Phys. Rev. D \textbf{65}, 084014 (2002); R. G. Cai
and Q. Guo \emph{ibid}. \textbf{69}, 104025 (2004).

\bibitem{Od1}  M. Cvetic, S. Nojiri, S. D. Odintsov, Nucl. Phys. \textbf{B628%
}, 295 (2002); S. Nojiri and S. D. Odintsov, Phys. Lett. B \textbf{521}, 87
(2001).

\bibitem{Deh1}  M. H. Dehghani, Phys. Rev. D \textbf{67}, 064017 (2003);
\emph{ibid}. \textbf{69}, 064024 (2004); \emph{ibid}. \textbf{70}, 064019
(2004).

\bibitem{DM}  M. H. Dehghani and R. B. Mann, Phys. Rev. D \textbf{72},
124006 (2005); M. H. Dehghani and S. H. Hendi, hep-th/0602069.

\bibitem{Deh2}  M. H. Dehghani and M. Shamirzaie, Phys. Rev. D \textbf{72},
124015 (2005).

\bibitem{Hoi}  F. Muller-Hoissen, Phys. Lett. B \textbf{163}, 106 (1985).

\bibitem{Mal}  J. Maldacena, Adv. Theor. Math. Phys., \textbf{2}, 231
(1988); E. Witten, \emph{ibid}. \textbf{2}, 253 (1998); O. Aharony, S. S.
Gubser, J. Maldacena, H. Ooguri and Y. Oz, Phys. Rept., \textbf{323}, 183
(2000).

\bibitem{Sken}  M. Hennigson and K. Skenderis, J. High Energy Phys. \textbf{7%
}, 023 (1998).

\bibitem{BK}  V. Balasubramanian and P. Kraus, Commun. Math. Phys. \textbf{%
208,} 413 (1999).

\bibitem{Od2}  S. Nojiri and S. D. Odintsov, Phys. Lett. B \textbf{444}, 92
(1998); S. Nojiri, S. D. Odintsov and S. Ogushi, Phys. Rev. D \textbf{62},
124002 (2000).

\bibitem{Deh3}  M. H. Dehghani, Phys. Rev. D \textbf{66}, 044006 (2002);
\emph{ibid}. \textbf{65}, 124002 (2002); M. H. Dehghani and A.
Khodam-Mohammadi, \emph{ibid}. \textbf{67}, 084006 (2003).

\bibitem{DM2}  M. H. Dehghani and R. B. Mann, Phys. Rev. D \textbf{64},
044003 (2001); M. H. Dehghani, \emph{ibid}. \textbf{65}, 104030 (2002); M.
H. Dehghani and H. KhajehAzad, Can. J. Phys. \textbf{81}, 1363 (2003).

\bibitem{dS}  A. Strominger, J. High Energy Phys. \textbf{10}, 034 (2001);
\textbf{11}, 049 (2001); V. Balasubramanian, P. Horova, and D. Minic, \emph{%
ibid.} \textbf{05}, 043 (2001); E. Witten, hep-th/0106109; D.
Klemm, Nucl. Phys. \textbf{B625}, 295 (2002); V. Balasubramanian,
J. deBoer, and D. Minic, Phys. Rev D \textbf{65}, 123508 (2002);
S. Nojiri and S. D. Odintsov, Phys. Lett. B \textbf{519}, 145
(2001); J. High Energy Phys. \textbf{12}, 033 (2001); Phys. Lett.
B \textbf{523}, 165 (2001); \textbf{528}, 169 (2002); R. G. Cai,
Phys. Lett. B \textbf{525}, 331 (2002); Nucl. Phys. \textbf{B628},
375 (2002); R. G. Cai, Y. S. Myung, and Y. Z. Zhang, Phys. Rev. D
\textbf{65}, 084019 (2002); R. Bousso, A. Maloney, and A.
Strominger, \emph{ibid.} \textbf{65}, 104039 (2002); A. M.
Ghezelbash and R. B. Mann, J. High Energy Phys. \textbf{01}, 005
(2002); M. H. Dehghani, Phys. Rev. D \textbf{65}, 104003 (2002).

\bibitem{Awad}  A. M. Awad, Class.Quant.Grav. \textbf{20}, 2827 (2003).

\bibitem{ATZ}  R. Aros, R. Troncoso and J. Zanelli, Phys. Rev. D \textbf{63}%
, 084015 (2001).

\bibitem{Haw1}  S. W. Hawking, Commun. Math. Phys. \textbf{25}, 152 (1972);
S. W. Hawking and G. F. R. Ellis, \emph{The Large Scale Structure of
Spacetime} (Cambridge University Press, 1973).

\bibitem{Beck}  J. D. Beckenstein, Phys. Rev. D \textbf{7}, 2333 (1973); S.
W. Hawking, Nature (London) \textbf{248}, 30 (1974); G. W. Gibbons and S. W.
Hawking, Phys. Rev. D \textbf{15}, 2738 (1977).

\bibitem{Haw2}  C. J. Hunter, Phys. Rev. D \textbf{59}, 024009 (1999); S. W.
Hawking, C. J. Hunter and D. N. Page, \emph{ibid}. \textbf{59}, 044033
(1999); R. B. Mann, \emph{ibid}. \textbf{60}, 104047 (1999); \emph{ibid}.
\textbf{61}, 084013 (2000);

\bibitem{fails}  M. Lu and M. B. Wise, Phys. Rev. D \textbf{47},
R3095,(1993); M. Visser, \emph{ibid}. \textbf{48}, 583 (1993).

\bibitem{Myers2}  T. Jacobson and R. C. Myers, Phys. Rev. Lett. \textbf{70},
3684 (1993); R. M. Wald, Phys. Rev. D \textbf{48}, R3427, (1993); M. Visser,
\emph{ibid}. \textbf{48}, 5697 (1993); T. Jacobson, G. Kang and R. C. Myers,
\emph{ibid}. \textbf{49}, 6587,(1994); V. Iyer and R. M. Wald, \emph{ibid}.
\textbf{50}, 846 (1994).

\bibitem{Ross}  T. Clunan, S.F. Ross and D. J. Smith, Class Quant. Grav.
\textbf{21}, 3447 (2004).

\bibitem{Gub}  M. Cvetic and S. S. Gubser, J. High Energy Phys. \textbf{04},
024 (1999); M. M. Caldarelli, G. Cognola and D. Klemm, Class. Quantum Grav.
\textbf{17}, 399 (2000).

\bibitem{Wit2}  E. Witten, Adv. Theor. Math. Phys. \textbf{2}, 505 (1998).
\end{thebibliography}
\end{document}